\documentclass[twocolumn,floatfix,showpacs,preprintnumbers,amsmath,amssymb,prb,superscriptaddress]{revtex4}
\usepackage{graphicx}
\usepackage{dcolumn}
\usepackage{bm}
\usepackage{subfigure}

\begin{document}
\title{N\'eel and Valence Bond Crystal Order on a 
  Distorted Kagom\'e Lattice: Implications For Zn-Paratacamite}
\author{Erik S. S{\o}rensen}
 \affiliation{Department of Physics and Astronomy, McMaster University, Hamilton, Ontario, Canada L8S 4M1}
\email{sorensen@mcmaster.ca}
\author{Michael J. Lawler}
 \affiliation{Department of Physics, Applied Physics and Astronomy, Binghamton University, PO Box 6000, Binghamton, NY 13902-6000}
 \affiliation{Department of Physics, Cornell University, Ithaca, NY 14853}
\author{Yong Baek Kim}
 \affiliation{Department of Physics, University of Toronto, Toronto, Ontario M5S 1A7, Canada}

 \date{\today}
\begin{abstract}
Zn-Paratacamite is a rare spin 1/2 antiferromagnetic insulator with an ideal
kagom\'e lattice structure in part of its phase diagram. As a function of Zn
doping, this material undergoes a structural distortion which relieves the
frustration and introduces magnetic order in the ground state, though the
precise nature of the order is not clear at this point.  In this paper, we
present strong evidence for N\'eel ordering in the {\it strongly} distorted phase of
Zn-Paratacamite through the application of quantum Monte-Carlo techniques.
These numerical results support a recent Schwinger-boson mean field
theory of Zn-Paratacamite. For {\it weak} distortion, close to the ideal kagom\'e limit,
our results indicate a regime with no N\'eel order but with a broken glide-plane symmetry.
For this model the glide-plane symmetry is broken by {\it any} valence bond crystal.
Hence, our results lend support to recent 
proposals\cite{Nikolic:2003hz,Singh:2007wv} of a
valence bond crystal ground state for the undistorted lattice. 
The phase
transition between the two phases could be in the deconfined universality class if it is
not a first order transition.
\end{abstract}


\maketitle
\section{Introduction}

Recently, a number of spin 1/2 frustrated magnetic insulators have been
discovered without any sign of magnetic order or structural distortions down to
the lowest temperatures studied\cite{Shimizu:2003eo, Hiroi:2001ej,
Helton:2007wx, Okamoto:2007fk}.  Among these materials, the Zn doped
Paratacamite family stands out for having a (nearly) controllable degree of
distortion allowing the amount of geometric frustration to be tuned directly
by an experimentalist. As such, they are a promising place to look for new
phases of matter while at the same time probe how these new phases may be
related to more well understood phases.

\begin{figure}
\includegraphics[width=0.4\textwidth]{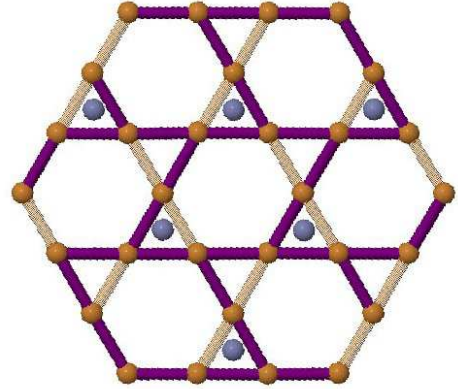}
\caption{(Color online) The layered Zn Paratacamite magnetic lattice structure. 
Cu atoms live on the (sometimes distorted) kagom\'e layer (bronze atoms) while Zn or Cu atoms occupy 
sites on a triangular lattice above the kagom\'e plane (grey atoms). 
Solid (purple) bonds represent nearest neighbors ($\langle ij\rangle$ in Eq. \eqref{eq:ham}) resulting in a `brick-wall' lattice, while transparent (bronze) bonds represent the next nearest neighbors ($\langle\langle ij\rangle\rangle$ in Eq. \eqref{eq:ham}). }
\label{fig:ZnPara}
\end{figure}

The control of the distortion is largely through the chemical pressure induced
by the substitution of Zn atoms for Cu atoms on the (grey) triangular lattice
planes that live in between kagom\'e-planes, as shown in Fig. \ref{fig:ZnPara}.
While Zn and Cu atoms are similar in size, Zn atoms fit into these sites
without disrupting their environment, unlike Cu atoms which distort the
kagom\'e plans given a high enough density.  In particular, for less than 0.3
filling of Zn atoms (greater than 0.7 filling of Cu atoms) the lattice
distorts in a remarkable bi-partite structure and magnetic order is found in
the ground state\cite{Lee:2007ly,Helton:2007wx,Kim:2008ko}. The spins are thus
relatively unfrustrated at these low doping concentrations. For Zn doping
larger than this threshold, the lattice has the undistorted ideal kagom\'e form
and for $x\ge 0.4$ no magnetic ordering has been reported down to 50 mK despite
an estimated spin exchange $J\sim 200 K$\cite{Lee:2007ly}.  

A natural theoretical model of this material is the spin 1/2 Heisenberg model
with two exchange parameters on a distorted kagom\'e lattice\cite{Lee:2007ly,
Lawler:08,Kim:2008ko} (see Fig. \ref{fig:ZnPara}):
\begin{equation}
H=J\sum_{\langle i,j\rangle}{\bf S}_i\cdot{\bf S}_j + \lambda_D J \sum_{\langle\langle ij\rangle\rangle} {\bf S}_i\cdot{\bf S}_j
\label{eq:ham}
\end{equation}
Here $\lambda_D$ tunes the distortion on the next nearest neighbor bonds (nnn)
with $\lambda_d=1$ the undistorted ideal kagom\'e limit (where nnn bonds are
equivalent to nn bonds). This is an idealized model for several reasons:
it assumes the coupling between planes and further neighbors is weak (which
seems reasonable\cite{Lee:2007ly,Kim:2008ko}), it neglects
Dzyaloshinsky-Moria interactions possibly important for the low temperature
susceptibility\cite{Rigol:2007pq, Tovar:2008df} , and it replaces the effect
of doping in the triangle lattice planes with the uniform distortion
parameter $\lambda_D$. While it is possible that any of these approximations
may be important for some properties of Zn Paratacamite, here we will focus
on those properties which clearly belong to the phenomenology of this
simplified model.

In this paper, we study the ground state properties of the Hamiltonian in Eq.
\eqref{eq:ham} as a function of $\lambda_D$, extrapolating between a bi-partite `brick-wall' lattice
at $\lambda_D=0$ and the isotropic kagom\'e lattice at $\lambda_D=1$.  
At $\lambda_D=0$, we show using valence bond quantum
Monte-Carlo\cite{Sandvik:05,Beach:06}, that the ground state is magnetically
ordered with the expected N\'eel pattern for this bi-partite lattice and with a
magnetization of $m^\dagger = 0.240(1)$ that is 22\% smaller than the square
lattice value\cite{Sandvik:08}. For $0\le \lambda_D\le 1$ we study this model using
exact diagonalization on finite size clusters of size 12, 24 and 36 sites. By
introducing symmetry breaking fields, we study the susceptibility of the ground
state towards dimerization. Remarkably, we find that for $\lambda_D\gtrsim
0.8$, a phase transition occurs towards a rotationally invariant state which
prefers to have a broken glide-plane symmetry, consistent with 
the presence of a VBC order including 
the pin-wheel VBC pattern proposed by Ref. \onlinecite{Lawler:08}.  This symmetry breaking survives
up to the $\lambda_D=1$ ideal kagom\'e limit. While it is difficult to draw
definitive conclusions on such small systems, a broken glide-plane symmetry support Refs.
\onlinecite{Nikolic:2003hz, Singh:2007wv,Singh:08} proposal that the spin 1/2 kagom\'e
antiferromagnet has a valence bond crystal (VBC) ground state. At the same time a broken glide-plane
symmetry is not consistent with a spin-liquid phase, frequently supported by other exact diagonalization studies\cite{Misguich:07}. In addition, while we
can't rule out a first order transition from a VBC phase to the N\'eel phase in
our model, it is also possible this quantum phase transition is in the
deconfined universality class\cite{Senthil:2004qw}.

\section{Results at $\lambda_D=0$}
We first discuss our results obtained at $\lambda_D=0$ where we have been able to study large system. As can be seen from Figs.~\ref{fig:ZnPara}, where
the transparent (bronze) bonds are proportional to $\lambda_D$, the lattice formed by the remaining solid (purple) bonds 
is a bi-partite `brick-wall' lattice with a coordination number
of 3 on two thirds of the sites and of 2 on the remaining sites. Due to the bi-partite nature of the $\lambda_D=0$ lattice there is no frustration.
A classical AF N\'eel state can be unambiguously assigned to the lattice. It is then possible to perform very efficient quantum Monte Carlo
simulations using the recently proposed~\cite{Sandvik:05,Beach:06} valence bond quantum Monte Carlo (VBQMC). For $\lambda_D=0$ there is no sign problem
and extremely precise results can be obtained. VBQMC is a projection method where the $T=0$ ground-state is projected out through
the repeated application of the hamiltonian, $H$, on a trial state, $|\Psi_T\rangle.$ In essence, $|\Psi_G\rangle=(-H)^n|\Psi_T\rangle.$
In the limit where $n\to\infty$ this becomes exact. In a practical implementation $n$ is kept fixed at a high number and the different
terms in $|\Psi_G\rangle$ are sampled using Monte Carlo methods. For convergence, the relevant lattice size independent expansion order is 
$n/N_b$ where $N_b$ is
the number of terms in the Hamiltonian. $N_b$ is equal to $(4/3)N$ for the brick-wall lattice with $N$ the number of sites in the lattice.
Typically we use $n/N_b=3-10$ and an extrapolation
to $n/N_b=\infty$ can then be performed.

\begin{figure}[tbp]
\begin{center}
\includegraphics[clip,width=8cm]{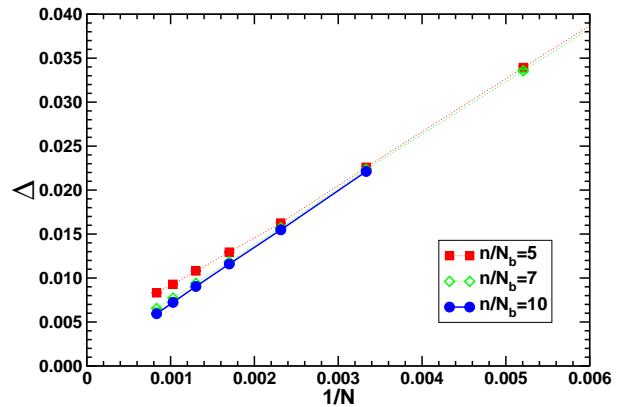}
  \caption{(Color online.) The singlet-triplet gap, $\Delta$, versus inverse system size $1/N$. 
The different curves correspond to different expansion orders $n$, with the ratio $n/N_b$
 kept fixed. Here, $N_b$ is the number of bond operators in the Hamiltonian, $N_b=(4/3) N$.
Results are shown for $n/N_b=5$ ($\blacksquare$), $n/N_b=7$ ($\diamond$), $n/N_b=10$ ($\bullet$).
Error bars are shown but are typically smaller than the symbols.}
  \label{fig:gap}
\end{center}
\end{figure}

We have performed systematic VBQMC studies of $\lambda_D=0$ brick-wall lattices with number of sites $N=12\times m^2$ for $m=1,\ldots 10$ using periodic 
boundary conditions. Typically, $10^6-10^7$ MCS were performed for a range of values of $n/N_b=3,5,7,10$. 
All errrorbars were calculated using standard binning techniques.

A very natural question to ask is if the $\lambda_D=0$ brick-wall lattice has a non-zero singlet to triplet gap, $\Delta$.
A particularly appealing feature of VBQMC is that it allows for a direct estimator~\cite{Sandvik:05,Beach:06} of this gap
independent of the estimators for the ground-state singlet and excited triplet energies. Due to a cancellation of errors
it is then possible to calculate this gap with a precision significantly exceeding that which could have been obtained by
separately calculating the ground and excited
state energies.
Our results for $\Delta$ at $\lambda_D=0$ are shown in Fig.~\ref{fig:gap}. Data are shown for 3 different values of
$n/N_b=5$ ($\blacksquare$), $n/N_b=7$ ($\diamond$), $n/N_b=10$ ($\bullet$) versus inverse system size $1/N$. 
At $N=1200,\ n/N_b=10$ the gap is $\Delta=0.0059(1)J$ and minimal dependence on the expansion
order $n/N_B$ is seen. From the results shown in Fig.~\ref{fig:gap} we conclude that 
the gap vanishes in the thermodynamic limit.

\begin{figure}[tbp]
\begin{center}
\includegraphics[clip,width=8cm]{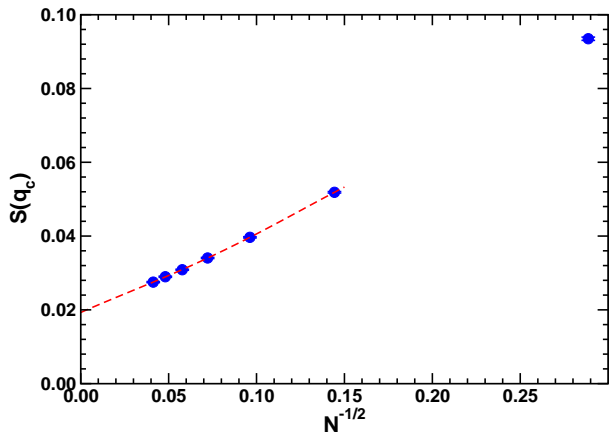}
  \caption{(Color online.) The structure factor, $S({\bf q}_c)$, versus inverse linear system size $1/\sqrt{N}$. 
Results are shown for $n/N_b=10$.
Error bars are shown but are typically smaller than the symbols. Results for $n/N_b=7$ are indistinguishable from
  the $n/N_b=10$ results shown and have been left out for clarity.}
  \label{fig:M2}
\end{center}
\end{figure}

For the two-dimensional square lattice anti-ferromagnetic Heisenberg model it is well known~\cite{Reger:88}
that the anti-ferromagnetic order exists at $T=0$ with $m^\dagger=0.30743(1)$~\cite{Sandvik:08}. The square lattice has a coordination 
number of 4 where as the brick-wall lattice has a mixed coordination of 2 and 3. We therefore expect $m^\dagger$
to be smaller or possibly zero for the brick-wall lattice.
As usual, we define:
\begin{equation}
S({\bf q}_c)=\frac{1}{N^2}\left\langle \left(\sum_{x,y}\tilde S^z(x,y)\right)^2\right\rangle,
\end{equation}
where ${\bf q}_c$ is the wave-vector of the staggered magnetization and $\tilde S^z(x,y)$ is given by:
\begin{equation}
\tilde S^z(x,y) = \frac{1}{2}\epsilon_{x,y}\sigma^z(x,y),
\end{equation}
with $\epsilon_{x,y}$ equal to $+1$ or $-1$ depending on what sublattice the point $(x,y)$ belongs to.
Hence we have~\cite{Reger:88}:
\begin{equation}
m^\dagger=\langle \tilde S^z_i\rangle =\lim_{L\to\infty}\sqrt{3S({\bf q}_c)}
\end{equation}
Our results for $S({\bf q_c})$ for $n/N_b=10$ are shown in Fig.~\ref{fig:M2}. 
It is expected~\cite{Reger:88} that 
the leading finite size corrections are of the form $1/\sqrt{N}$ and a fit to this form yields $S({\bf q_c})=0.0192(5)$
and consequently:
\begin{equation}
m^\dagger=0.240(3).
\end{equation}
As expected, this value is reduced with respect to the square lattice result, but is clearly non-zero, indicating
a well established anti-ferromagnetic order at $\lambda_D=0$.

\section{Results at $\lambda_D\neq0$}
We now turn to a discussion of our results for $0<\lambda_D\leq 1$. In this case it is no longer possible to perform VBQMC calculations
due to a sign problem that appears rather severe as soon as $\lambda_D\neq0$ and reliable numerical results are therefore much harder to obtain. 
In light of the strong sign problem we have performed exact diagonalization studies for $0<\lambda \leq 1$ on
finite size systems employing periodic boundary conditions. Our goal is to study generalized bond susceptibilities with respect to symmetry breaking fields.
\begin{figure}[tbp]
\begin{center}
\includegraphics[clip,width=4cm]{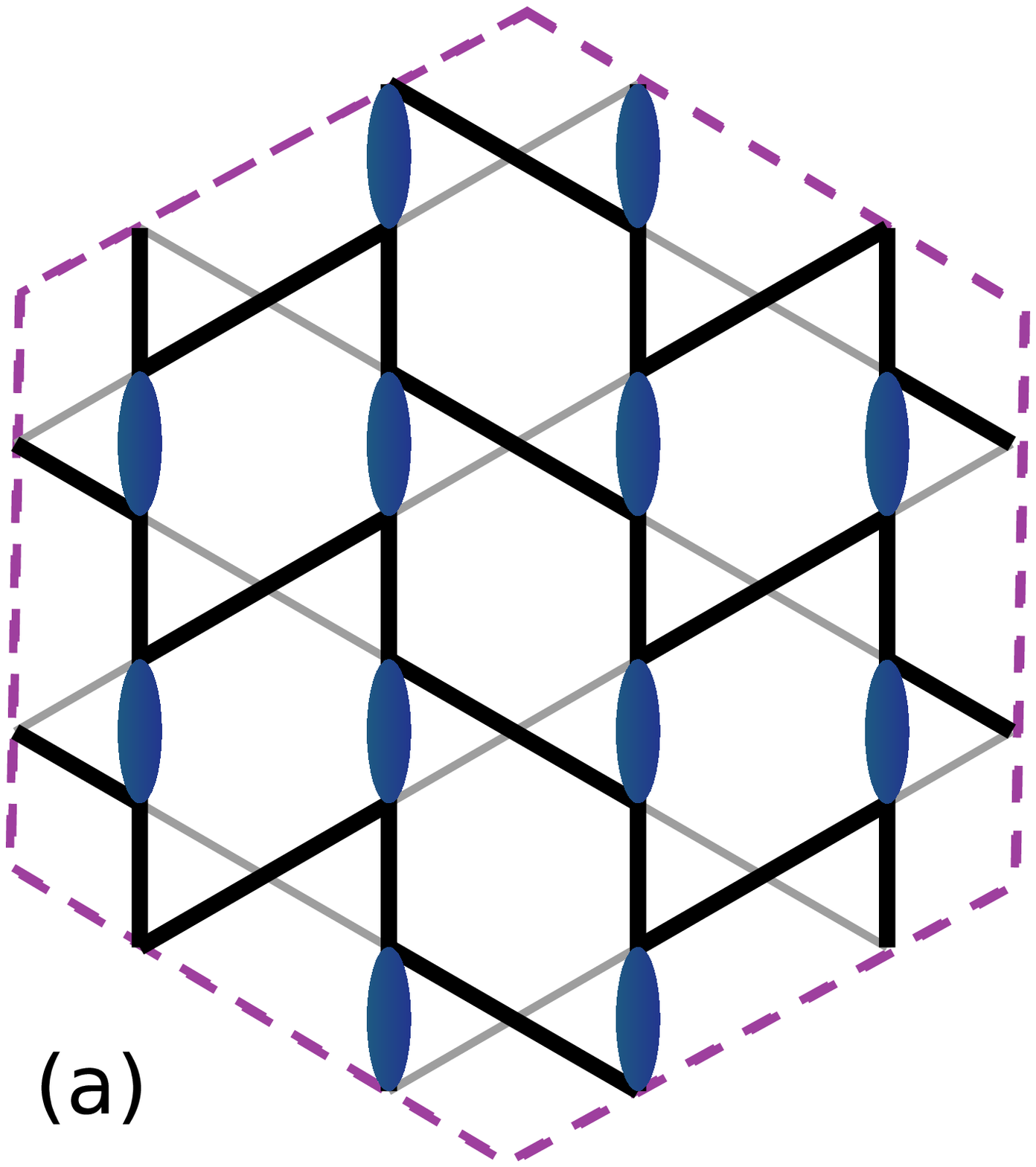}
\includegraphics[clip,width=4cm]{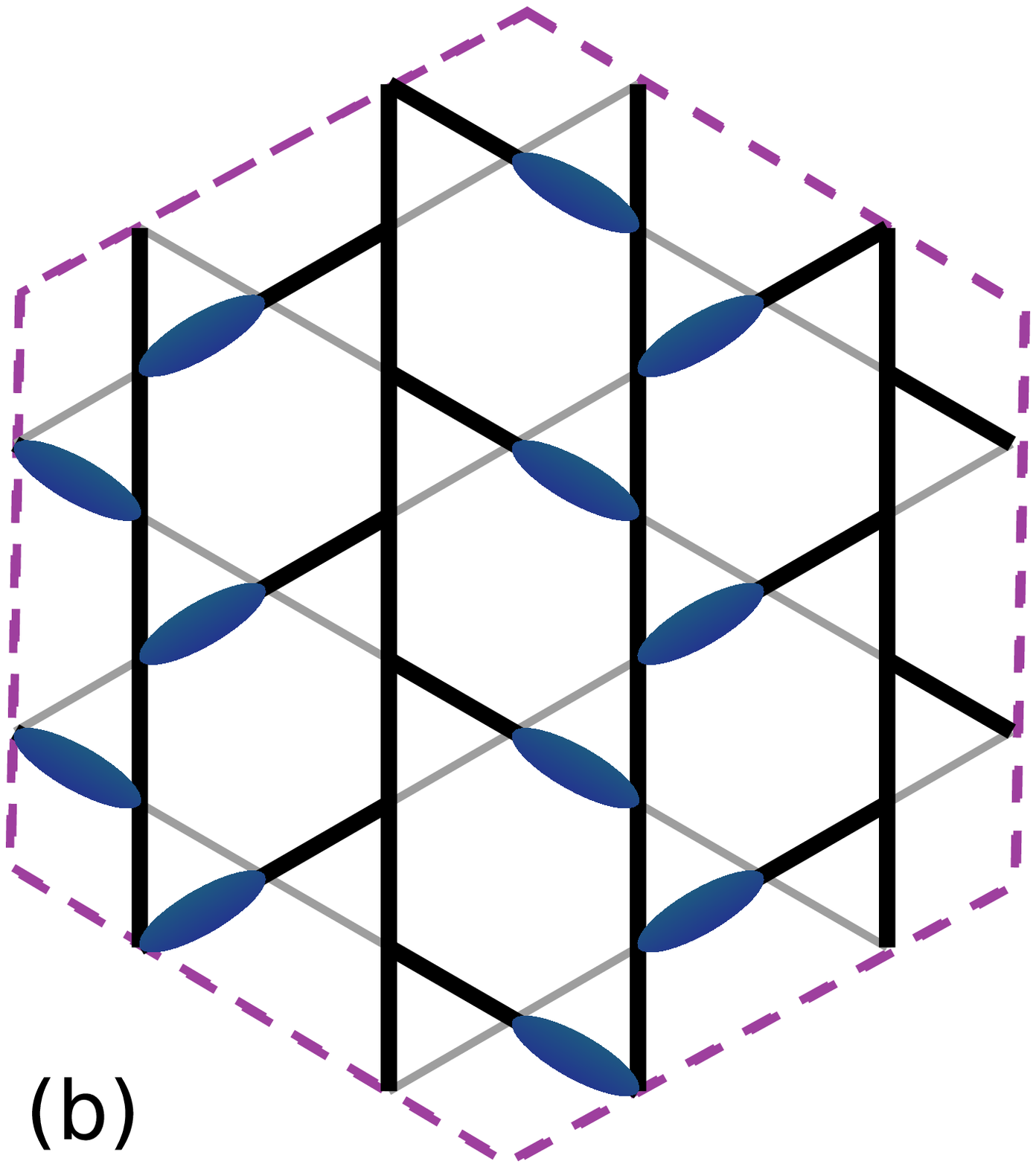}
  \caption{(Color online.) (a) The glide-plane symmetry breaking field. (b) $C_2$ symmetry breaking field. Solid dimers denote bonds
  where the coupling strength is modified to $J'$.}
  \label{fig:fields}
\end{center}
\end{figure}
We focus on $C_2$ and glide-plane (GP) symmetry breaking fields shown in Fig.~\ref{fig:fields} where the dimers indicate bonds where the strength
is modified $J'=J\pm\delta$. The $C_2$ symmetry corresponds to a rotation by $\pi$ and clearly the field shown in Fig.~\ref{fig:fields}(a) breaks this
symmetry. The GP symmetry~\cite{Lawler:08} is somewhat more exotic and corresponds to a translation along the rails where the dimers are sitting followed by a reflection
around one of these rails. We note that the GP field does not break the $C_2$ symmetry and likewise the $C_2$ field preserves the GP symmetry. The pin-wheel
VBC discussed in Ref.~\onlinecite{Lawler:08} would break the GP symmetry but not the $C_2$ symmetry where as the columnar VBC~\cite{Lawler:08} would break both.
If we by ${\bf b,b_{C_2,{\rm GP}}}$
denote the ground-state expectation value $\langle{\bf S}_i\cdot {\bf S}_j\rangle$ for the bond ${\bf b}$ and its partner under the symmetry operation ${\bf b_{C_2,{\rm GP}}}$, we can define
the generalized bond susceptibility as follows:
\begin{eqnarray}
&&\chi_{C_2,{\rm GP}}=\nonumber\\
&&\ \lim_{\delta\to 0}\frac{|\Delta {\bf b}_{C_2,{\rm GP}}(J'=J+\delta)-\Delta {\bf b}_{C_2,{\rm GP}}(J'=J-\delta)|}{2\delta},\nonumber\\
  \label{eq:chi}
\end{eqnarray}
with
\begin{equation}
\Delta {\bf b}_{C_2,{\rm GP}}={\bf b}(J')-{\bf b}_{C_2,{\rm GP}}(J).
\end{equation}
Clearly, if $\Delta {\bf b}$ goes to zero linearly with $\delta$ the generalized bond susceptibility is a constant independent of system size and
the associated symmetry is not spontaneously broken. On the other hand, a bond susceptibility diverging with system size signals that the associated
symmetry is spontaneously broken in the thermodynamic limit.

\begin{figure}[tbp]
\begin{center}
\includegraphics[clip,width=5cm]{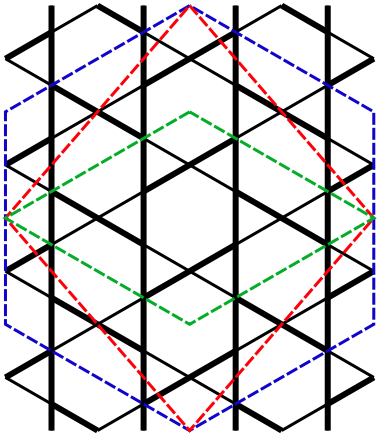}
  \caption{(Color online.) The $12$, $24$ and $36$ site lattices.}
  \label{fig:unitcells}
\end{center}
\end{figure}

When performing exact diagonalization studies of small systems the choice of the finite cluster is crucial since the smaller clusters will
reduce the point group symmetry of the infinite lattice. For the isotropic kagom\'e lattice, $\lambda_D=1$, the plane group is $p6mm$.
This symmetry group implies that for the isotropic kagom\'e lattice all bonds are equivalent. Our choice of finite clusters are shown in Fig.~\ref{fig:unitcells}
for $N=12,24,36$. Only the $N=36$ cluster has the full symmetry point group symmetry of the kagom\'e lattice. However, for all clusters do we find that all
bonds are equivalent at $\lambda_D=1$. For $0<\lambda_D<1$ only two different type of bonds occur for these clusters. These are the only clusters we have found
with these properties. For the bond susceptibilities to yield meaningful information about the thermodynamic limit this is very important since we want to
make sure that the presence of a reduced point group symmetry doesn't explicitly break the $C_2$ or GP symmetry. This is not the case for the clusters shown in Fig.~\ref{fig:unitcells}.

\begin{figure}[tbp]
\begin{center}
\includegraphics[clip,width=8cm]{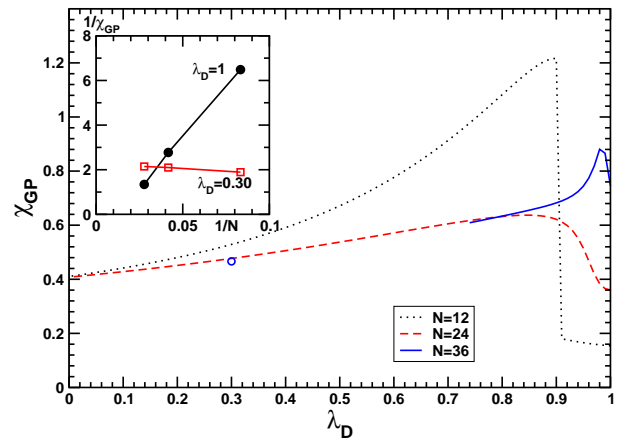}
  \caption{(Color online.) The glide-plane susceptibility, $\chi_{\rm GP}$ as a function of $\lambda_D$ for the different
    system sizes, $N=12,24,36$. The circles represent results for $N=36, \lambda_D=0.3$. The inset shows $\chi_{\rm GP}^{-1}$ versus $1/{N}$ at $\lambda_D=0.3,1.0$.}
  \label{fig:ChiGP}
\end{center}
\end{figure}

\begin{figure}[tbp]
\begin{center}
\includegraphics[clip,width=8cm]{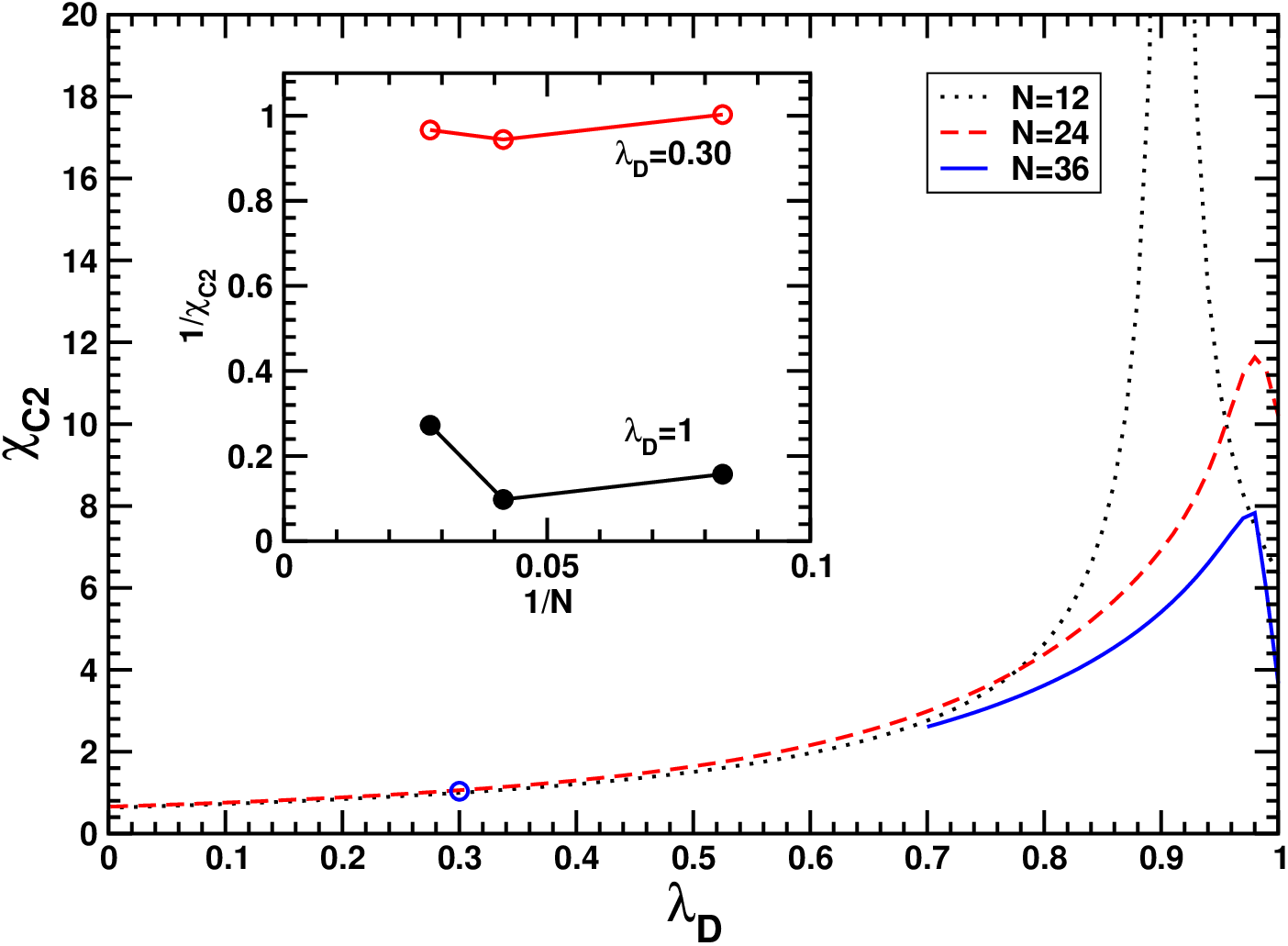}
  \caption{(Color online.) The C$_2$ susceptibility, $\chi_{\rm C2}$ as a function of $\lambda_D$ for the different
    system sizes, $N=12,24,36$. The circles represent results for $N=36, \lambda_D=0.3$. The inset shows $\chi_{\rm C2}$ versus $1/{N}$ at $\lambda_D=0.3,1.0$.}
  \label{fig:ChiC2}
\end{center}
\end{figure}

Using the clusters from Fig.~\ref{fig:unitcells} we can now study $\chi_{\rm GP}$ and $\chi_{C_2}$ as a function of $\lambda_D$ for the
different clusters. Our results are shown in Figs~\ref{fig:ChiGP},\ref{fig:ChiC2} respectively. We have used $\delta\leq 0.001$ small enough
that $\chi$, Eq.~(\ref{eq:chi}), is almost completely independent of $\delta$. We begin by discussing $\chi_{\rm GP}$ shown in Fig.~\ref{fig:ChiGP}.
For $\lambda_D$ less than roughly $\sim 0.8$ do we find that $\chi_{\rm GP}$ is almost independent of $N$.
In the inset is shown $1/\chi_{\rm GP}$ as a function of $1/N$ for $\lambda=0.3$ indicating a finite value in the thermodynamic limit.
This is consistent with the GP symmetry not being broken. However, for $\lambda_D$ greater than $\sim 0.8$ pronounced size dependence occurs. 
At $\lambda_D=1$, $1/\chi_{\rm GP}$ as a function of $1/N$ is shown in the inset. In this case it seems reasonable to conclude that $\chi_{\rm GP}$
diverges with $N$ and hence that the GP symmetry is spontaneously broken in the thermodynamic limit. A natural interpretation of this result
is that a quantum phase transition occurs at $\sim  0.8$ between a state with anti-ferromagnetic order, that does not break the GP symmetry, 
to a {\it new} phase  where the GP symmetry is {\it broken}.

Finally, in Fig.~\ref{fig:ChiC2} we show our results for $\chi_{C_2}$. Again we see that for $\lambda_D$ smaller than roughly $\sim 0.8$ there is
very little $N$ dependence. In the inset in Fig.~\ref{fig:ChiC2} is shown $1/\chi_{C_2}$ as a function of $1/N$ at $\lambda_D=0.3$. Clearly the
results extrapolate to a finite value in the thermodynamic limit consistent with the absence of $C_2$ symmetry breaking as would be the case
for an anti-ferromagnetic phase.
As before, we find that for $\lambda_D$ greater than roughly $\sim 0.8$ pronounced finite size effects occur consistent with a quantum phase transition.
However, in this case, as can be seen in the inset in Fig.~\ref{fig:ChiC2} at $\lambda_D=1$, the susceptibility {\it doesn't diverge} but rather
tends to a finite, possibly very small value in the thermodynamic limit. Therefore, in the new phase occurring for $\lambda_D$ greater than $~0.8$ 
the $C_2$ symmetry is not broken. 

\section{Discussion}
Given these results, we may draw several conclusions.
\begin{figure}
\subfigure[MZ pattern]{\includegraphics[width=0.2\textwidth]{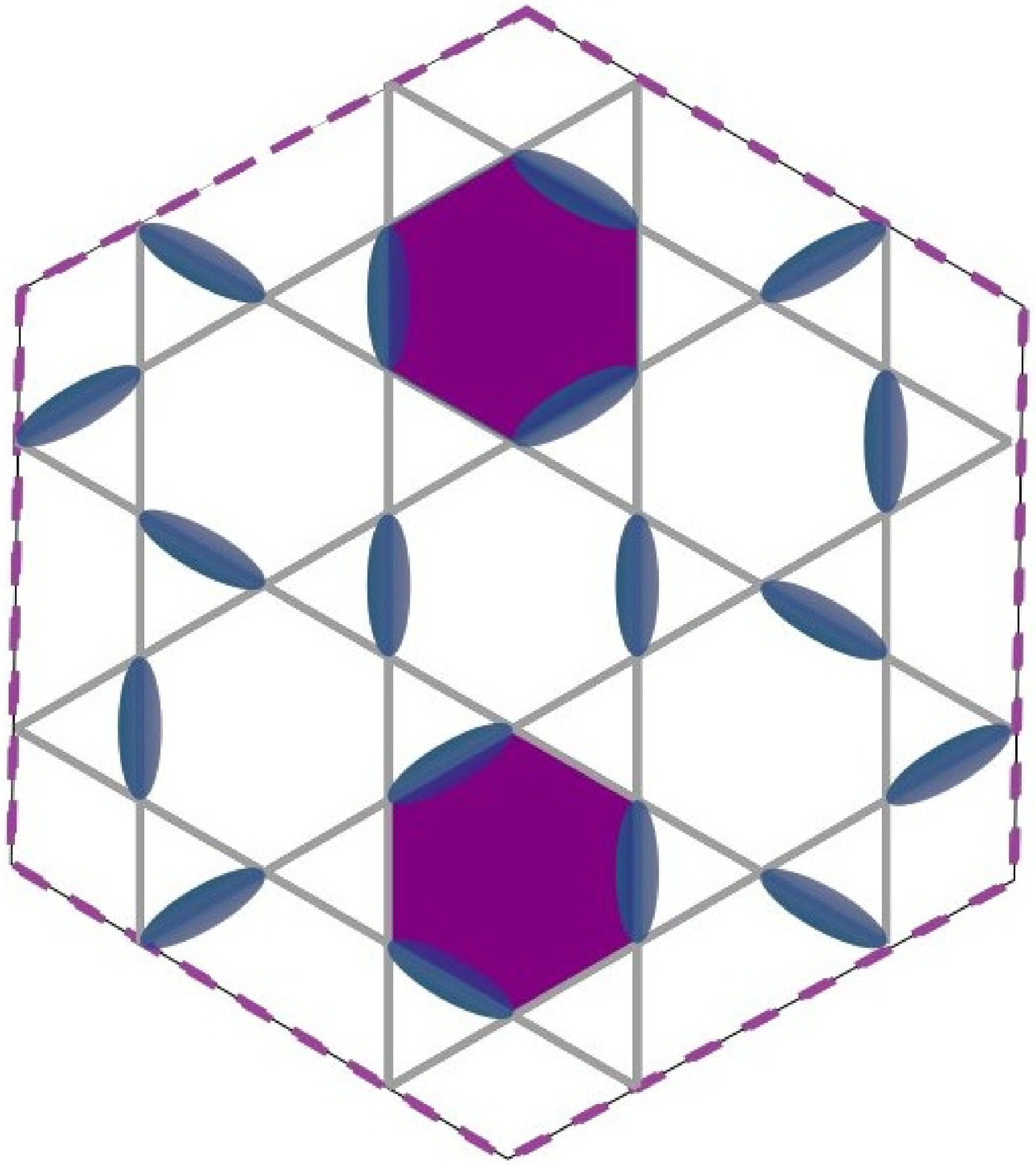}}
\subfigure[Pin-wheel pattern]{\includegraphics[width=0.2\textwidth]{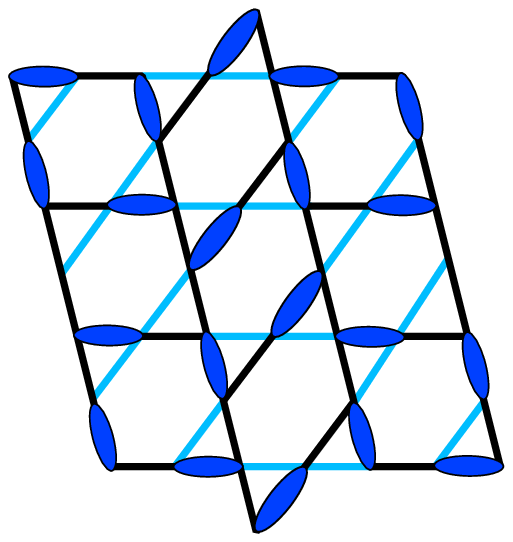}}
\caption{(a) The Marston and Zeng(MZ) dimer pattern~\cite{MZ:91} on the kagom\'e lattice and (b) the pin-wheel state on the distorted kagome lattice\cite{Lawler:08}. 
The MZ pattern arises by maximizing the number of dimers around each hexagon. 
Notice how the dimers on the two highlighted hexagons can be rotated by 180 about the center of the hexagon without needing to alter the rest of the dimer pattern. This benzene-like resonance suggests that rotational symmetry may be only very weakly broken if this is the ground state. The pin-wheel pattern, on the other hand, maximizes dimers around each rhombus and is manifestly C$_2$ rotationally symmetric about each hexagon.}
\label{fig:MZ}
\end{figure}
At $\lambda_D=0$, we have demonstrated that the spin gap vanishes and that the
ground state has a finite staggered magnetization that is 22\% smaller than the
square lattice value. We find this surprising given that each site has either
two or three neighbors (with 2.67 neighbors on average) and that this network
is not far from the one dimensional chain model which is disordered. It appears
that this form of dimensional reduction does not easily suppress magnetic
order. 

While we could only study very small system sizes for $\lambda_D>0$, finite
size effects seem to be small all the way out to $\lambda_D\approx 0.8$. As a
result, the anti-ferromagnetic order is quite robust and appears to be the
ground state with a large basin of stability. 

The ground state for $0.8\lesssim\lambda_D\le1$ appears to break the glide
plane symmetry while remaining invariant under $C_2$ rotations. 
While larger systems would be required to make definitive
conclusions, this evidence is in stark contrast to the prediction that the
kagom\'e lattice ground state is a spin liquid~\cite{MZ:91,Waldtmann:98}. 

Remarkably, on this lattice all valence bond crystals break the glide plane
symmetry (the most glide-plane symmetric configuration of Fig.
\ref{fig:fields}(b) still breaks glide plane symmetry if the missing dimers
are added to the picture).  So the breaking of glide plane symmetry
strongly supports recent proposals\cite{Nikolic:2003hz,Singh:2007wv,Singh:08}
that the ground state maybe a valence bond crystal (VBC). Candidate such states
include the Marston and Zeng  36 site unit cell dimer (spin singlet)
pattern(MZ)~\cite{MZ:91} and the pin-wheel pattern (in the presence of
distortion) (see Fig.~\ref{fig:MZ}).  One may argue that the MZ pattern
should also break C$_2$ rotational symmetry. However, such a symmetry may
naturally be restored by benzene-like resonances on the three dimer hexagons.
One should note that whether the recent ED results~\cite{Misguich:07} are disfavoring the
MZ VBC state as well as other proposed VBC states~\cite{Syro:02,Nikolic:2003hz}
or not, is a subject of intense debate.\cite{Singh:08}

Since any VBC will have a diverging glide-plane susceptibility the results
presented here are not very sensitive to transitions between different VBC's as
long as they conserve $C_2$ symmetry. For example, a transition from the MZ
pattern at $\lambda_D=1$ to the pin-wheel pattern at $\lambda_D<1$ is certainly
possible.  For $0.8 \lesssim \lambda_D$ we can therefore not exclude the
presence of several different $C_2$ symmetric VBC phases although the rate of
divergence of $\chi_{GP}$ could potentially be quite different for different
phases. In fact, one might speculate that the cusp in $\chi_{GP}$ for $N=36$
and in $\chi_{C2}$ for $N=24,36$ in both cases at $\lambda_D^c=0.98$ is a signature
of a phase-transition between different valence bond crystals.

A phase transition near $\lambda_D\approx 0.8$ was also found in the large-N
study of Ref. \onlinecite{Lawler:08}. Thus both large-N and exact diagonalization
methods predict the existence of a quantum phases transition at a value of
$\lambda_D$ away from the ideal kagom\'e limit. If we assume that the spin gap
is non-zero at $\lambda_D=1$ and 
vanish approximately linearly with the deviation of $\lambda_D$ from 1, then 
this value for the quantum critical point is also roughly consistent
with the vanishing of the spin gap (which, from exact diagonalizations, is estimated to
rather small but finite~\cite{Lecheminant:97,Waldtmann:98} in the thermodynamic limit and
has a value of ~0.1848J at
 $\lambda_D=1$ for the 36-site cluster). Given the apparent glide plane symmetry breaking for
$\lambda_D\gtrsim 0.8$, this phase transition then appears to be between two
phases with {\it unrelated orders}. It may therefore fall into the deconfined
universality class~\cite{Senthil:2004qw} if it is not a first order transition. 

\begin{acknowledgments}
We thank Young Lee and Seung-Hun Lee for useful discussions. 
This work was supported by the Natural Sciences and Engineering
Research Council of Canada (ESS and YBK), the Canadian Foundation for Innovation (ESS),
the Canadian Institute for Advanced Research and the Canada Research Chair Program (YBK).
ESS gratefully acknowledge the hospitality of the department of physics at the university of Toronto 
where part of this work was carried out.
This work was made possible by the facilities of the Shared Hierarchical Academic Research 
Computing Network (SHARCNET:www.sharcnet.ca).
\end{acknowledgments}

\bibliography{qvb}

\end{document}